\documentclass[aps,prd,11pt,notitlepage,nofootinbib,superscriptaddress,showkeys,showpacs]{revtex4-2}
\pdfoutput=1
\usepackage{amstext,amsmath,amssymb,amsfonts,bbm,euscript,color,dsfont}
\usepackage[latin1]{inputenc}
\usepackage[english]{babel}
\usepackage{epsfig}
\usepackage{hyperref}
\usepackage{amsthm}
\usepackage{subfigure}
\usepackage{color}
\usepackage{multirow}
\usepackage{bbold}
\usepackage{tikz}
\usetikzlibrary{arrows,backgrounds}
\usepackage{verbatim} 
\usepackage{graphicx}
\usepackage{latexsym}
\usepackage[shortlabels]{enumitem}

\begin{document}

\title{Aspects of the gauge boson-gaugino mixing in a supersymmetric scenario with Lorentz-symmetry violation}
	
\author{R.~C.~Terin} \email{rodrigoterin3003@gmail.com}
\affiliation{Instituto Tecnol\'ogico de Aeron\' autica, 12.228-900 S\~ao Jos\'e dos Campos, SP, Brasil}

\author{W.~Spalenza}\email{wspalenza@gmail.com}
\affiliation{Instituto Federal do Esp\'{i}rito Santo,
29150-410 Cariacica, ES, Brasil}
\affiliation{Universidade Federal do Esp\'{i}rito Santo, 29075-910 Vit\'{o}ria, ES, Brasil}

\author{H.~Belich}\email{belichjr@gmail.com}
\affiliation{Universidade Federal do Esp\'{i}rito Santo, 29075-910 Vit\'{o}ria, ES, Brasil}

\author{J.~A.~Helay\"{e}l-Neto} \email{helayel@cbpf.br}
\affiliation{Centro Brasileiro de Pesquisas F\'{i}sicas, 22290-180 Rio de Janeiro, RJ, Brasil}

\begin{abstract}
We write down an $\mathcal{N}=1$ supersymmetric extension for non-Abelian
gauge theories in $(1+3)$ dimensions with a Lorentz- and CPT-violating term of
the Carroll-Field-Jackiw type. By including effects of the background
(supersymmetric) fermion bilinears that accompany Lorentz-symmetry violation
in a Carroll-Field-Jackiw scenario, we investigate both the gauge boson and
gaugino dispersion relations in order to compute their respective masses in
terms of the background structures. Such results open up a potential path
towards a possible mechanism for gaugino-gauge boson conversion, an analogue
of the Primakoff effect, induced here not by an external magnetic field, but
instead by the (Majorana) fermionic sector of the supersymmetry multiplet in
the backstage of the Lorentz-symmetry violation.

\end{abstract}
\maketitle

\affiliation{Instituto Tecnol\'ogico de Aeron\' autica, 12.228-900 S\~ao Jos\'e dos Campos, SP, Brasil}

\affiliation{Instituto Federal do Esp\'{i}rito Santo,
29150-410 Cariacica, ES, Brasil} \affiliation{Universidade Federal do Esp\'{i}rito Santo, 29075-910 Vit\'{o}ria, ES, Brasil}

\affiliation{Universidade Federal do Esp\'{i}rito Santo, 29075-910 Vit\'{o}ria, ES, Brasil}

\affiliation{Centro Brasileiro de Pesquisas F\'{i}sicas, 22290-180 Rio de Janeiro, RJ, Brasil}

\section{Introduction}

It is well-known that both Lorentz and charge-parity-time reversal (CPT)
symmetries are fundamental invariances in the formulation of the Standard
Model for Particle Physics (SM). However, at energies close to the Planck
scale, these symmetries may be violated \cite{Carroll:1989vb}; especially, the phenomenon of
Lorentz symmetry violation (LSV) may have its origin in a more
fundamental set-up, like superstring theory where supersymmetry (SUSY) plays an
important role
\cite{Kostelecky:1988zi,Kostelecky:1989jw,Kostelecky:1989jp,Kostelecky:1990pe,Kostelecky:1991ak,Kostelecky:1995qk,Kostelecky:2000hz,Kostelecky:1999mu}%
. At this more fundamental stage, SUSY might be either exact or a broken
symmetry, with the latter assumption still being an open issue. To our sense,
in the high-energy scale where Lorentz-symmetry breaking (LSV) effects first
appear, SUSY should still be exact or, if it is broken, its breaking scale is
not far enough from the LSV scale, so that supersymmetric imprints should not
be neglected in inspecting physics in a scenario with Lorentz-symmetry
violation
\cite{Berger:2001rm,Belich:2003fa,BaetaScarpelli:2003kx,GrootNibbelink:2004za,Bolokhov:2005cj,Feng:2005ba,Colladay:2010tx,Farias:2012ed,Belich:2013rma,Belich:2015qxa,Bonetti:2017toa}%
.

It is mandatory to quote formulations that have been proposed by incorporating
background tensor fields in the framework one refers to as the Standard Model
Extension (SME) \cite{Colladay:1996iz,Colladay:1998fq}. Specific formulations
widely investigated and discussed \cite{Colladay:1996iz,Colladay:1998fq},
adopt either a four-vector through a CPT-odd term \cite{Carroll:1989vb} or a
CPT-preserving term
\cite{KM1,KM2,Bailey, PRD2,Paulo,Klink2, Klink3,Kostelec,Kob, beto,
beto1}, which is, in this case, traceless, $\kappa_{F\,\mu\nu}^{\mu\nu} = 0$. The idea of including and keeping
SUSY preserved, while Lorentz symmetry is violated, has been earlier developed
in the work of Ref. \cite{Berger:2001rm}. Experiments have been elaborated to
investigate LSV physics \cite{data, 2}; thereby, limits and constraints can be
imposed on the parameters that signal deviations from the Lorentz symmetry in
the sector of electrons, neutrinos, photons and hadrons
\cite{Kostelecky:2006ta,Kostelecky:2001mb,Kostelecky:2002hh}. As for the
category of non-Abelian Lorentz-violating models, some works may be
found
\cite{Kostelecky:2018yfa,Kostelecky:2019fse,Santos:2014lfa,Santos:2016uds}
(and references therein); however, there is still some room left for a more
complete and systematic study of Yang-Mills models in connection with LSV,
since experimental constraints on the parameters of non-Abelian models are
still not really exhaustive \cite{Colladay:2006rk}.

Hence, our aim in this work is twofold: (i) we propose an extension of the
path adopted to connect SUSY and LSV according to the works
\cite{Belich:2003fa,BaetaScarpelli:2003kx,Belich:2013rma}, but now replacing
the photonic sector by a non-Abelian gauge model; our contribution actually
sets out to build up a super-Yang-Mills-Caroll-Field-Jackiw (SYM-CFJ) action
including an LSV term at the classical level; (ii) from the
supersymmetry-displaying action built up in step (i), we compute the gauge
boson and gaugino dispersion relations and find their respective masses in
terms of the parameters present in the supersymmetric background that
accomplishes LSV.

For these purposes, we assume SUSY is present from the on-set using a
superfield that accommodates the LSV background, and SUSY algebra is here not
modified; as a result, in our formulation, we shall conclude that, whenever
LSV takes place, SUSY is contemporarily spontaneously broken as it is disclosed
by the background fermion, as we shall show further on. It also comes out
that a set of background fermion bilinears emerge in the process of breaking
Lorentz symmetry. Strictly speaking, we intend to find a way to describe the
microscopic origin of the vectors and tensors that usually appear in the
description of the LSV background. In this context, we shall introduce a
constant background four-vector into the super-Yang-Mills action by coupling
it through a CFJ like-term while preserving gauge symmetry; however, as we
have already anticipated, SUSY will also be (spontaneously) broken, as it
shall be duly discussed later on.

Finally, we highlight that gauge symmetry is kept throughout the whole procedure we are going to present. LSV takes place in a scenario where SUSY is
present, and the breaking of Lorentz symmetry in the supersymmetric background
will naturally induce SUSY breaking, as it shall be shown in Section III. For
our proposal in this paper, we follow, as mentioned before, the original idea
of the Abelian case implemented in the work of Ref.\cite{Belich:2003fa}, in
order to set up the minimal supersymmetric extension for the CFJ-like term in
its non-Abelian version \cite{Colladay:2006rk}, i.e.
\begin{align}
\mathcal{S}_{CPT-odd}  &  =  \int\,d^{4}x\,\bigg[-\frac{1}{2}\,\epsilon
^{\mu\nu\kappa\lambda}v_{\mu}A_{\lambda}^{a}\partial_{\kappa}A_{\lambda}%
^{a}+\frac{g}{6}\epsilon^{\mu\nu\kappa\lambda}f^{abc}v_{\mu}A_{\nu}%
^{a}A_{\kappa}^{b}A_{\lambda}^{c}\bigg]\,,\label{CS}%
\end{align}
where $v_{\mu}$ is a background field and $A^{a}_{\nu}$ is the non-Abelian
gauge field. Also, we adopt that $v_{\mu}$ is a constant vector in the active sense, whose role is to break Lorentz symmetry in the CFJ way. Moreover, by using an appropriate superfield to describe the background, the model can be made free from higher spin excitations and, as a result of the supersymmetric structure, there will naturally arise self-couplings for the non-Abelian gauge fields, as it should be, and a mass-type mixing between the gauge boson and the gaugino fields induced by the fermion present in the supersymmetric background.

The organization of our paper follows the structure below. In Section
\ref{II}, we present a quick review on the $\mathcal{N}=1$ supersymmetric
non-Abelian model, with some specific comments concerning the formalism
adopted here. Next, in Section \ref{III}, we build up, at the classical level,
the $\mathcal{N}=1$-super-Yang-Mills-Caroll-Field-Jackiw action with Lorentz
symmetry violation expressed in component fields, starting from the
superspace approach with SUSY
and LSV already broken. After that, in Section \ref{gauge fielddis}, and, as a consequence from the background fermion
bilinears, we compute dispersion relations for both the gauge boson and the
gaugino, compute their respective masses in terms of the background parameters
and carry out a discussion of a Gordon-like decomposition of the gaugino
currents. Finally, in Section \ref{V}, we display our Concluding Comments and
Future Perspectives. In addition, in Appendix \ref{A}, we present the notation and
conventions adopted throughout our algebraic manipulations; in Appendix \ref{B}, we cast the details for the attainment of the gaugino mass spectrum.

\section{Warm-up with the N=1-supersymmetric non-Abelian Yang-Mills model}

\label{II}

The $\mathcal{N}= 1$ supersymmetric non-Abelian model shall be briefly
reviewed is this Section. $G$ is a simple non-Abelian gauge group with
generators $T^{a}$ that fulfill the algebra
\begin{equation}
\lbrack T^{a},T^{b}]=if^{abc}T^{c}.
\end{equation}
These generators are Hermitian, $(T^{a})^{\dagger}=T^{a}$, $f^{abc}$ are structure constants, and the set of
indices ($a,b,c...$) stand for the adjoint representation of G. As a second step, we present the chiral superfields (the matter sector) transforming under the following transformations 
\begin{equation}
\Phi_{i}^{\prime}  =\exp{(i\Xi^{a}T^{a})_{ij}}\Phi_{j}\,,  \qquad \bar{\Phi}_{i}^{\prime}    = \bar{\Phi}_{j} \exp{(-i\Xi^{a}T^{a})_{ji}}\label{25}%
\end{equation}
and ($\Xi^{a}$, $\bar{\Xi^{a}}$) are both Lorentz-scalar
superfields that accommodate the (local) parameter of the gauge transformations and  which, respectively, obey the chirality and anti-chirality constraints in superspace,
\begin{align}
\bar{\mathcal{D}}_{\dot{\alpha}}\Xi = 0\,, \qquad \mathcal{D}_{\alpha}\bar{\Xi}  &  =0\,,
\label{500}
\end{align}
for all $a=1,2,...,n$; the SUSY covariant derivatives are defined in Appendix A, eq. \ref{cov_deriv_alpha}. Notice that, in eq.\eqref{500}, the bar operation, conventionally adopted to act on the the scalar superfields, corresponds to the complex conjugation operation:
\begin{equation}
    \bar{\Xi}^{a} = \Xi^{\star\,a}\,.
\end{equation}
The matter kinetic term, $\int d^{4}xd^{4}\theta
\,\bar{\Phi}_{i}\Phi_{i}$ is no longer invariant under \eqref{25},
actually, it is transformed into
\begin{align}
\int d^{4}xd^{4}\theta\,\bar{\Phi}\exp{(-i\bar{\Xi})}\exp{(i\Xi)}\Phi\,,
\label{510}
\end{align}
where the exponential factors above do not cancel each other because $\bar{\Xi}$ and $\Xi$ are different matrices in view of eq.\eqref{510}; $\bar{\Phi}$ is defined in the same way as $\bar{\Xi}$, namely, the complex conjugate of $\Phi$. In order to restore the invariance of $\bar{\Phi}\Phi$, let us introduce the
compensating factor $\exp(V)$, where $V\equiv V^{a}(x,\theta,\bar{\theta
})T^{a}$, the $V^{a}$'s being superfields which transform under $G$ as
follows:
\begin{eqnarray}
\exp{(V^{\prime})}=\exp{(i\bar{\Xi)}}\exp{(V)}\exp{(-i\Xi)}\,, \label{transf 2}%
\end{eqnarray}
therefore, the new term
\begin{align}
\mathcal{S}_{matter-gauge}  & = \int d^{4}xd^{4}\theta\,\bar{\Phi}\exp{(gV)}\Phi\,,
\end{align}
is invariant under the transformations \eqref{25} and \eqref{transf 2} and the
usual kinetic terms for matter fields carry out their interactions with the
gauge fields. Also, we must mention that $g$ is the gauge coupling constant. Having established an action for matter-gauge fields, the next
step is to obtain the supersymmetric action for the pure Yang-Mills sector.
The strategy is similar to the conventional non-Abelian gauge theories: one
defines gauge covariant derivatives and, by considering their commutators, one
attains the field strengths which will, later on, be used to define the pure
gauge action.

The supersymmetric covariant derivatives shall be indicated, in a
superspace formulation, by
\begin{eqnarray}
\nabla_{A} &=&(\partial_{\mu};\mathcal{D}_{\alpha},\bar{\mathcal{D}}_{\dot{\alpha}})\,
\end{eqnarray}
where
\begin{eqnarray}
\left\{ \mathcal{D}_{\alpha},\bar{\mathcal{D}}_{\dot{\beta}}\right\}  &=& -2i\sigma^{\mu
}_{\alpha\dot{\beta}}\partial_{\mu}\,.
\end{eqnarray}
and their respective gauge covariantizations are established if we impose that they transform according to
\begin{align}
\nabla_{A}^{\prime}  & = \exp{(i\bar{\Xi})}\nabla_{A}\exp{(-i\Xi)}\,.
\end{align}
This yields the expression $\bigg(\exp{(-gV)}\mathcal{D}_{\alpha} \exp{(gV)}\bigg)$ to be the gauge covariant version of the spinorial superspace derivative, from which we can read off the gauge connection superfield:

\begin{equation}
\exp{(-gV)}\mathcal{D}_{\alpha} \exp{(gV)} \quad\longrightarrow\quad
\exp{(i\bar{\Xi})}\bigg(\exp{(-gV)}\mathcal{D}_{\alpha} \exp{(gV)}
\bigg)\exp{(-i\Xi)}\,,
\end{equation}
thus, we can finally write down the gauge covariant supersymmetric derivative as follows
\begin{equation}
\mathcal{D}_{\alpha}^{\prime}  = \exp{(-gV)}\mathcal{D}_{\alpha} \exp{(gV)} =
\mathcal{D}_{\alpha} + \Gamma_{\alpha}\,,\\
\end{equation}
where the gauge potential superfield turns out to be
\begin{equation}
\Gamma_{\alpha}  = \exp{(-gV)}\bigg(\mathcal{D}_{\alpha} \exp{(gV)}%
\bigg) \,.
\end{equation}

In order to present the super-Yang-Mills kinetic term, we need to first define
the following fundamental supersymmetric field strength:
\begin{equation}
W_{\alpha}\equiv\bar{\mathcal{D}}^{2}\Gamma_{\alpha}= \bar{\mathcal{D}}
^{2}\Bigg[\exp{(-gV)}\bigg(\mathcal{D}_{\alpha} \exp{(gV)} \bigg)\Bigg]\,,
\end{equation}
for which the chirality condition is satisfied, namely, $\mathcal{D}\bar
{W}=\bar{\mathcal{D}}W=0$. The Lie-algebra-valued vector superfield, $V$, is
Hermitean, $V=V^{\dagger}$. This superfield can be written in the Wess-Zumino
gauge as
\begin{equation}
V_{WZ}(x,\theta,\bar{\theta})=\theta^{\alpha}\sigma^{\mu}_{\alpha\dot{\alpha}%
}\bar{\theta}^{\dot{\alpha}}A_{\mu} (x)+\theta^{2}\bar{\theta}_{\dot{\alpha}%
}\bar{\lambda}^{\dot{\alpha}}(x)+\bar{\theta}^{2} \theta^{\alpha}%
\lambda_{\alpha} (x)+\theta^{2}\bar{\theta}^{2}D (x). \label{vct-supf}%
\end{equation}
Here, we have defined the gauge field as $A_{\mu}$, the Majorana spinors
$(\lambda,\bar{\lambda})$ and the auxiliary field $D$. Moreover, the gauge
transformation acting on the field-strength superfield strength reads as
below:
\begin{equation}
W_{\alpha}^{\prime}=\exp{(-i\Xi)}\,W_{\alpha}\exp{(i\Xi)},
\label{transf 4}%
\end{equation}
and, similarly, for the Hermitian conjugate. Now, by expanding in a power
series using eq. (\ref{vct-supf}), one sees
\begin{equation}
W^{a}_{\alpha}=\frac{g}{2}\bar{\mathcal{D}}^{2}\mathcal{D}_{\alpha}\left(
V^{a}_{WZ}+igf^{abc}V^{b}_{WZ}V^{c}_{WZ}\right) ,
\end{equation}
with $V_{WZ}^{2}=\frac{1}{2}\theta^{2}\bar{\theta}^{2}A^{\mu}A_{\mu},\,$
$V_{WZ}^{3}=0$ (we have omitted the group index for a moment). Indeed, to build up the action for the gauge superfield, we simply take the square of $W^{\alpha}$. As it is a chiral (spinor) superfield, we write
\begin{equation}
\mathcal{S}_{SYM}=-\frac{1}{64g^{2}}\,Tr\,\int d^{4}xd^{2}\theta
\,\bigg[\bigg(\,W^{\alpha}W_{\alpha}+ c.c\bigg) \bigg]\,, \label{ga}%
\end{equation}
where $c.c.$ denotes the complex conjugate term. We have a more familiar action
written in terms of component fields as shown below:
\begin{equation}
\mathcal{S}_{SYM}=\int d^{4}x\,\left(  -\frac{1}{4}F^{\mu\nu\,a}F_{\mu\nu}^{a}
-\frac{i}{2}\lambda^{\,a}\sigma^{\mu}\mathcal{D}_{\mu}^{ab}\bar{\lambda}%
^{\,b}-\frac{i}{2}\bar{\lambda}^{\,a}\bar{\sigma}^{\mu}\mathcal{D}_{\mu}%
^{ab}{\lambda}^{\,b}+\frac{1}{2}\left(  D^{a}\right) ^{2} \right)
\,.\label{12}%
\end{equation}
Here, $Tr(T^{a}T^{b})=k\delta^{ab}$ and, in our case, we shall take $k= 1$. We
are denoting the Yang-Mills field strength and the fermionic covariant
derivative in the adjoint representation according to which is given below:
\begin{align}
F_{\mu\nu}^{a}(x)  &  =\partial_{\mu}A_{\nu}^{a}-\partial_{\nu}A_{\mu}
^{a}-gf^{abc}A_{\mu}^{b}A_{\nu}^{c},\label{fs}\\
\mathcal{D}_{\mu}^{ab}  &  =\delta^{ab}\partial_{\mu}-gf^{abc}A_{\mu}^{c}.
\end{align}

Once these main elements have been presented, in the next section we shall be
working out the $\mathcal{N}=1$-supersymmetric version of the CFJ model by
following a similar procedure, as mentioned previously, adopted in the
formulation of the Abelian model
\cite{Belich:2003fa,BaetaScarpelli:2003kx,Belich:2013rma}.


\section{The $\mathcal{N}=1$-supersymmetric Yang-Mills-Carroll-Field-Jackiw
model}

\label{III}

We shall now be presenting our super-Yang-Mills-Carroll-Field-Jackiw (SYM-CFJ)
model. We start by considering the gauge transformation of the quantity
below:
\begin{equation}
W^{\prime\,\alpha}\Gamma_{\alpha}^{\prime}=\exp{(-i\Xi)}\bigg(W^{\alpha}%
\mathcal{D}_{\alpha}\bigg)\,\exp{(i\Xi)}+ \exp{(-i\Xi)}\,\bigg(W^{\alpha}\Gamma_{\alpha
}\bigg)\,\exp{(i\Xi)}.
\end{equation}
Here, we are neglecting surface terms and $\mathcal{D}\bar{\Xi}%
=\bar{\mathcal{D}}\Xi=0.$ In four space-time dimensions, SUSY invariance
is ensured by the following term: $Tr\,\left(  W^{\alpha}\Gamma_{\alpha
}S+c.c\right)  $. Therefore, the gauge and SUSY invariant action for the
Carroll-Field-Jackiw model reads as follows:%

\begin{equation}
\mathcal{S}_{CPT-odd}^{(SUSY)}= \frac{1}{g}\int d^{4}xd^{2}\theta d^{2}%
\bar{\theta}\,\bigg( \,\,W^{\alpha\, a}\Gamma_{\alpha}^{a}S+\bar{W}%
_{\dot{\alpha}}^{a}\bar{\Gamma}^{\dot{\alpha}\,a}\bar{S}\bigg) , \label{lb}%
\end{equation}
where $S$ is the scalar background superfield and $\Gamma_{\alpha}^{a}$ is described in adjoint representation as
\begin{align}
\Gamma_{\alpha}^{a}  &  =  \sigma_{\alpha\dot{\alpha}}^{\mu}\bar{\theta}%
^{\dot{\alpha}}A_{\mu}^{a}+2\theta_{\alpha}\bar{\theta}_{\dot{\alpha}}%
\bar{\lambda}^{a\dot{\alpha}}+\bar{\theta}^{2}\lambda_{\alpha}^{a}+\bar
{\theta}^{2}\bigg[2\delta_{\alpha}^{\beta}D^{a}+(\sigma^{\mu\nu})^{\beta
}_{\alpha}F_{\mu\nu}^{a}\bigg]\theta_{\beta}\nonumber\\
&  -  \frac{i}{2}\theta^{2}\sigma_{\alpha\dot{\alpha}}^{\mu}\varepsilon
^{\dot{\alpha}\dot{\beta}}\bar{\theta}\bar{\theta}\partial_{\mu}\bar{\lambda
}_{\dot{\beta}}^{a}\,.
\end{align}
Note that $S$ is invariant under $G$, i.e
$\,\delta_{G}S=0$ with the following expansion
\begin{align}
S(x,\theta,\bar{\theta})  &  =s(x) +i\theta^{{\alpha}}\sigma^{\mu}_{\alpha
\dot\alpha}\bar{\theta}^{\dot\alpha}\partial_{\mu}s(x) -\frac{1}{4}\theta
^{2}\bar{\theta}^{2}\Box s(x) +\sqrt{2}\theta^{\alpha}\psi_{\alpha
}(x)\nonumber\\
&  +\frac{i}{\sqrt{2}}\theta^{2}\bar{\theta}_{\dot\alpha}\bar{\sigma}%
^{\mu{\dot\alpha}\alpha}\partial_{\mu}\psi_{\alpha}(x) +\theta^{2}F(x)\,.
\label{bcg-supf}%
\end{align}
This superfield is neutral under the gauge group and its canonical mass
dimension is $1$. Since this background superfield is a chiral
one, the maximum spin component of the background is $(\frac{1}{2})$,
characterizing the supersymmetric partner of a dimensionless scalar field.
Moreover, eq. \eqref{bcg-supf} is invariant under the gauge group
transformations, $S^{\prime}=S.$ Thus, our total action,
$\eqref{12}+\eqref{lb}$, can be projected out in terms of the superfield components by carrying out the superspace integrations and, finally, reads as below:
\begin{align}
\mathcal{S}_{CPT-odd}^{(SYM)}  &  =  \int d^{4}x\,\bigg\{-\bigg[\frac{1}%
{4}+\frac{(s+s^{*})}{2}\bigg]F_{\mu\nu}^{a}F^{\mu\nu\,a}+i\,\varepsilon
^{\mu\nu\kappa\lambda}\,\partial_{\mu}\left( s-s^{*}\right) A_{\nu}%
^{a}\partial_{\kappa}A_{\lambda}^{a}\nonumber\\
&  -  \frac{ig}{3}\varepsilon^{\lambda\nu\kappa\mu}f^{abc}\,\partial_{\mu
}\left( s-s^{*}\right) A_{\nu}^{a}A_{\kappa}^{b}A_{\lambda}^{c}+\bigg[\frac
{1}{2}+4\left( s+s^{*}\right) \bigg]D^{a}D^{a}\nonumber\\
&  -  \bigg[\frac{1}{2}-2s\bigg]\left( i\lambda^{a}\sigma^{\mu}\mathcal{D}%
_{\mu}^{ab}\bar{\lambda}^{b}\right) -\bigg[\frac{1}{2}-2s^{*}\bigg]\left(
\,i\bar{\lambda}^{a}\bar{\sigma}^{\mu}\mathcal{D}_{\mu}^{ab}\lambda^{b}\right)
-\sqrt{2}\left( \lambda^{a}\sigma^{\mu\nu}\psi\right) F_{\mu\nu}%
^{a}\nonumber\\
&  +  \sqrt{2}\left( \bar{\lambda}^{a}\bar{\sigma}^{\mu\nu}\bar{\psi}\right)
F_{\mu\nu}^{a}+(\lambda^{a})^{2}F+(\bar{\lambda}^{a})^{2}F^{*}-2\sqrt{2}%
\bar{\lambda}^{a}\bar{\psi}D^{a}-2\sqrt{2}\lambda^{a}\psi D^{a}%
\bigg\}\,,\label{SuCS}%
\end{align}
with the field strength tensor defined by eq. \eqref{fs}. At this point, we should clarify to the reader that the CFJ-term comes naturally out with the the vector $v_{\mu}$ given by the four-gradient of the complex component, s, of the superfield S. This means that it is SUSY the responsible for this specific form of $v_{\mu}$. Without SUSY, the only requirement on $v_{\mu}$ is that its curl must vanish, so as to ensure that gauge invariance is preserved SUSY specifies $v_{\mu}$ one step further: by associating it with a gradient, the curl is automatically zero so that gauge symmetry is guaranteed. Now, from eq. \eqref{SuCS},
we highlight a remark already observed in the study of the Abelian case
\cite{Belich:2003fa}. One notices that, whenever $s+s^{*}=0$, there remains the non-trivial imaginary part of $s$, responsible for the appearance of the vector $v_{\mu}$, which implements Lorentz-symmetry breaking in the CFJ way.

As we have already identified the background field, $v_{\mu}$, let us analyse
SUSY's transformations for the background superfield $S$:
\begin{align}
\delta s  & = \sqrt{2}\epsilon\psi\,,\nonumber\\
\delta\psi_{\alpha} &  = \sqrt{2}F\epsilon+ i\sqrt{2}\sigma_{\alpha\dot
{\alpha}}^{\mu}\bar{\epsilon}^{\dot{\alpha}}\partial_{\mu}s\,,\nonumber\\
\delta F &  = i\sqrt{2}\bar{\epsilon}_{\dot{\alpha}}\sigma^{\mu\,\dot{\alpha
}\alpha}\partial_{\mu}\psi_{\alpha}\,.
\end{align}
One required condition for $v_{\mu}$ being constant is $\partial_{\mu}s\neq0$.
Hence for the configuration type $S=(\partial_{\mu}s\neq0,\psi=0,F=0)$, one
sees the unavoidable breaking of SUSY. In this sense, we understand that SUSY
breaking is naturally induced by LSV. In our viewpoint $\delta\psi\neq0$
because of $\partial_{\mu}s\neq0$, thus $\partial_{\mu}s\equiv v_{\mu}$
defines a privileged direction in space-time and this anisotropy which breaks
Lorentz symmetry leads to the generation of the Goldstone fermionic mode
$(\partial_{\mu}\psi\neq0)$. As a result, there takes place the spontaneous
supersymmetry breaking (SSB) phenomenon. Also, in order to compute both gauge
boson and gaugino dispersion relations in the next Section, we shall rewrite
the action \eqref{SuCS} in terms of Majorana's 4-component spinors, i.e.,
$\Lambda^{a} =
\begin{pmatrix}
\lambda^{a}_{\alpha}\\
\bar{\lambda}^{a}_{\dot{\alpha}}%
\end{pmatrix}
$ in charge of gauginos's description and $\Psi=
\begin{pmatrix}
\psi\\
\bar{\psi}%
\end{pmatrix}
$ representing the fermionic component of $S$. Furthermore, by using
Euler-Lagrange's equations, we can remove the auxiliary field, $D$, and, by
Fierzing the spinor bilinears, we finally have the following (on-shell)
action:
\begin{align}
\mathcal{S}_{CPT-odd}^{(SYM)}  &  =  \int d^{4}x\,\bigg\{-\frac{1}{4}F_{\mu
\nu}^{a}F^{\mu\nu\,a}-\frac{1}{2}\,\varepsilon^{\mu\nu\kappa\lambda}\,v_{\mu
}A_{\nu}^{a}\partial_{\kappa}A_{\lambda}^{a}-\frac{g}{6}\varepsilon^{\mu
\nu\kappa\lambda}f^{abc}\,v_{\mu}A_{\nu}^{a}A_{\kappa}^{b}A_{\lambda}%
^{c}\nonumber\\
&  -  \frac{i}{2}\bar{\Lambda}^{a}\gamma^{\mu}\mathcal{D}_{\mu}^{ab}%
\Lambda^{b}+\bigg(Re(F)+\frac{1}{4}\bar{\Psi}\Psi\bigg)\bar{\Lambda}%
^{a}\Lambda^{a}-i\bigg(Im(F)+\frac{1}{4}i\bar{\Psi}\gamma_{5}\Psi
\bigg)\bar{\Lambda}^{a}\gamma_{5}\Lambda^{a}\nonumber\\
&  -  \frac{1}{4}\left( \,v_{\mu}+\bar{\Psi}\gamma_{\mu}\gamma_{5}\Psi\right)
\bigg(\bar{\Lambda}^{a}\gamma^{\mu}\gamma_{5}\Lambda^{a}\bigg)+\sqrt{2}%
\bar{\Lambda}^{a}\Sigma^{\mu\nu}\gamma_{5}\Psi F_{\mu\nu}^{a}%
\bigg\}\,,\label{SuCFJ}%
\end{align}
with $\Sigma^{\mu\nu}\equiv\frac{i}{4}[\gamma^{\mu},\gamma^{\nu}]$. From
eq. \eqref{SuCFJ}, one can readily observe some important aspects; for instance,
we could be able to find naturally the term presented in eq.\eqref{CS}. In
addition, the appearance of bilinear terms created by the background spinor,
$\Psi$, which, along with the scalar $F$ yield massive terms for the gaugino
sector. We rename them as
\begin{align}
M_{1} & \equiv Re(F)+\frac{1}{4}\bar{\Psi}\Psi\,,\nonumber\\
M_{2}  & \equiv Im(F)+\frac{i}{4}\bar{\Psi}\gamma_{5}\Psi\,,\nonumber\\
R_{\mu}  & \equiv v_{\mu}+\bar{\Psi}\gamma_{\mu}\gamma_{5}\Psi\,.\label{130}%
\end{align}
Also, we shall point out the existence of self-couplings for the non-Abelian gauge sector, i.e. in the last term of eq.\eqref{SuCFJ} the fermionic background field, $\Psi$, triggers the coupling of the gauge bosons, using the associated gauge field strength, to the gauginos \footnote{This kind of coupling is also possible
via scalar background field $s$ as one can see in the action \eqref{SuCS}.}.
This special term borrows an interesting similarity with the Primakoff effect \cite{PhysRev.81.899}, whose phenomenology is based on the axion-photon conversion in presence of a very strong external magnetic field. Let us recall that the axion is considered to be an important sector of the dark matter fraction of our Universe \cite{Sikivie:2020zpn}. According to our description, the background fermion, $\Psi$, present in the superfield $S$ plays the role of the external magnetic field and the decay rate of (massive) gauginos into (massive) gauge bosons comes out proportional to the $\Psi$-condensate. Also, this term may be regarded as a non-Abelian version of the Primakoff effect since it displays a three-vertex gauge coupling of the gaugino to a pair of gauge fields, as described by the last term of the component-field action above. This three-vertex induces a one-loop radiative correction, proportional to the $\Psi$-condensate and the square of the gauge coupling constant, is generated to shift the tree-level gaugino mass. The latter, as we shall see in the incoming Section, appears given by the modulus of the space-component of the four-vector $v_{\mu}$. On the other hand, from this same three-vertex, one may compute the decay rate of the gaugino into a gauge boson pair. This process is energetically possible for, as we shall compute in the next Section, the gaugino and gauge boson masses are compatible with such a decay. We shall comment on that in our Concluding Comments. Finally, we also point out that energy-momentum conservation is ensured since the SUSY background components are space-time homogeneous and, then, no explicit dependence on the space-time coordinates is present in the action \eqref{SuCFJ}.



\section{The gauge boson and gaugino dispersion relations}

\label{gauge fielddis}

In this Section, in order to obtain the gauge bosons and gauginos dispersion
relations and their corresponding masses, we must compute, as a first step,
their field equations. For the gauge bosons, we have the
following equation:
\begin{align}
\mathcal{D}_{\mu}^{ab}F^{\mu\nu\,b}+v_{\mu}\tilde{F}{}^{\mu\nu\,a}  &  =
-2\sqrt{2}\bar{\Psi}\Sigma^{\mu\nu}\gamma_{5}\bigg(\mathcal{D}_{\mu}%
^{ab}\Lambda^{b}\bigg)\,.\label{gauge field0}%
\end{align}
Let us split the field strength in terms of the Yang-Mills electric and
magnetic fields, namely $(\vec{E}^{a}, \vec{B}^{a})$, with the following
definitions
\begin{align}
E_{i}^{a}=F_{0i}^{a}\,,  &  \qquad B_{i}^{a}=-\frac{1}{2}\varepsilon
_{ijk}F_{jk}^{a}\nonumber\\
\vec{E}^{a}=-\vec{\nabla}\Phi^{a}-\frac{\partial\vec{A}^{a}}{\partial
t}+gf^{abc}(\Phi^{b}\vec{A}^{c})\,,  &  \qquad \vec{B}^{a}=\vec{\nabla}%
\times\vec{A}^{a}-\frac{g}{2}f^{abc}\bigg(\vec{A}^{b}\times\vec{A}%
^{c}\bigg)\,,\label{33}%
\end{align}
and therefore eq.\eqref{gauge field0} can be decomposed in four classical
Yang-Mills equations given in what follows below.

\begin{itemize}

\item The Gauss-Yang-Mills law for the electric sector:
\end{itemize}%

\begin{align}
\label{40}\nabla\cdot\vec{E}^{a}+gf^{abc}\,\vec{A}^{b}\cdot\vec{E}^{c}
-\vec{v}\cdot\vec{B}^{a} &  =  -i\sqrt{2}\bar{\Psi}\gamma^{0}\gamma_{5}%
\vec{\gamma}\cdot\left( \nabla\Lambda^{a}+g f^{abc}\vec{A}^{b}\Lambda
^{c}\right) \,.\nonumber\\
\end{align}

\begin{itemize}
\item The Gauss-Yang-Mills law for the magnetic sector:
\end{itemize}%

\begin{align}
\nabla\cdot\vec{B}^{a}+gf^{abc}\,\vec{A}^{b}\cdot\vec{B}^{c}  &  =
0\,.\label{41}%
\end{align}

\begin{itemize}
\item The Faraday-Lenz-Yang-Mills law:
\end{itemize}%

\begin{align}
\nabla\times\vec{E}^{a}+ gf^{abc}\,\vec{A}^{b}\times\vec{E}^{c}  &  =
-\frac{\partial\vec{B}^{a}}{\partial t}+gf^{abc}\,\Phi^{b}\vec{B}%
^{c}\,.\label{42}%
\end{align}

\begin{itemize}
\item The Amp\`{e}re-Maxwell-Yang-Mills law:
\end{itemize}%

\begin{align}
\nabla\times\vec{B}^{a}+gf^{abc}\,\vec{A}^{b}\times\vec{B}^{c} -v_{0}\vec
{B}^{a} +\vec{v}\times\vec{E}^{a}  & = \frac{\partial\vec{E}^{a}}{\partial t}
+ gf^{abc}\,\Phi^{b}\vec{E}^{c}\nonumber\\
& +i\sqrt{2}\bar{\Psi}\gamma^{0}\gamma_{5}\vec{\gamma}\left( \frac
{\partial\Lambda^{a}}{\partial t} +gf^{abc}\,\Phi^{b}\Lambda^{c}\right)
\nonumber\\
& +\sqrt{2}\bar{\Psi}\gamma^{0}\vec{\gamma}\times\left( \nabla\Lambda^{a}
+gf^{abc}\,\vec{A}^{b}\Lambda^{c}\right) .\label{43}%
\end{align}

In addition, the Dirac equation for the gaugino field is established as
\begin{align}
i\gamma^{\mu}\mathcal{D}_{\mu}^{ab}\Lambda^{b}-2M_{1}\Lambda^{a}+2iM_{2}%
\gamma_{5}\Lambda^{a}+\frac{1}{2}R_{\mu}\gamma^{\mu}\gamma_{5}\Lambda
^{a}-\sqrt{2}\Sigma^{\mu\nu}\gamma_{5}\Psi F_{\mu\nu}^{a}  &  =
0\,.\label{gauginos1}%
\end{align}

To obtain the mass spectrum of the gauge bosons and gauginos, and
verify the SUSY break, let's consider the free fields and we will only iterate
the equations in linearized form. Thus, we consider solutions in plane waves
for the fields $\vec{E}^{a}$ and $\Lambda^{a}$.%

\begin{align}
\omega\vec{v}\times\vec{E}_{0}^{a}  &  =  -i\omega^{2}\vec{E}_{0}^{a}-\sqrt
{2}\omega^{2}\bigg(\bar{\Psi}\gamma_{0}\vec{\gamma}\gamma_{5}\Lambda_{0}%
^{a}\bigg)\,\nonumber\\
\bigg(\omega\delta_{ij}-i\varepsilon_{ijk}v_{k}\bigg)E_{0j}^{a}  &  =
-i\sqrt{2}\omega\bigg(\bar{\Psi}\gamma_{0}\gamma_{i}\gamma_{5}\Lambda_{0}%
^{a}\bigg)\,,
\end{align}
by calling $\Omega_{ij}=\omega\delta_{ij}-i\varepsilon_{ijk}v_{k}$, one has
the following equation
\begin{align}
\Omega_{ij}E_{0j}^{a}  &  =  -i\sqrt{2}\omega\bigg(\bar{\Psi}\gamma_{0}%
\gamma_{i}\gamma_{5}\Lambda_{0}^{a}\bigg)\,.
\end{align}

After that, we go along the same lines as it is done in the case of an Abelian
gauge field: to study the classical mass spectrum, we switch off the non-linear
terms and consider the plane wave solutions, since we are, for this purpose,
bound to the linearization regime. We then rewrite eq.\eqref{gauginos1} in
terms of the non-Abelian electric and magnetic fields, which yields:
\begin{align}
& \gamma^{\mu}k_{\mu}\Lambda_{0}^{a}-2M_{1}\Lambda_{0}^{a}+2iM_{2}\gamma
_{5}\Lambda_{0}^{a}+\frac{1}{2}R_{\mu}\gamma^{\mu}\gamma_{5}\Lambda_{0}%
^{a}\nonumber\\
&  =  -i\,\sqrt{2}\gamma^{0}\gamma_{5}\gamma^{i}\Psi\,E_{0i}^{a}-\sqrt{2}%
\gamma^{0}\gamma^{i}\Psi\, B_{0i}^{a}\,.\label{47}%
\end{align}
From eq. \eqref{47} with $\vec{k}=\vec{0}$, one has
\begin{align}
&   \omega\gamma^{0}\Lambda_{0}^{a}+2M_{1}\Lambda_{0}^{a}+2iM_{2}\gamma
_{5}\Lambda_{0}^{a}+\frac{1}{2}R^{0}\gamma^{0}\gamma_{5}\Lambda_{0}^{a}%
-\frac{1}{2}\vec{R}\,\cdot\vec{\gamma}\gamma_{5}\Lambda_{0}^{a}\nonumber\\
&  =  +i\,\omega\sqrt{2}\gamma^{0}\gamma_{5}\gamma_{i}\Psi\,\bigg[-i\sqrt
{2}\Omega_{ij}^{-1}\bigg(\bar{\Psi}\gamma_{0}\gamma_{j}\gamma_{5}\Lambda
_{0}^{a}\bigg)\bigg]\,,\label{50}%
\end{align}
where we have considered that the Faraday-Lenz-Yang-Mills magnetic field, at the linearized regime, may be expressed as $B_{0i}^{a} = \frac{1}{\omega}(\vec{k}\times\vec{E}_{0}^{a})_{i}$ and
\begin{align}
\bigg(\Omega_{ij}\bigg)^{-1}  &  =  \frac{\omega}{\omega^{2}-v^{2}}\delta
_{ij}-\frac{i}{\omega^{2}-v^{2}}\varepsilon_{ijk}v_{k}-\frac{1}{\omega
\bigg(\omega^{2}-v^{2}\bigg)}v_{i}v_{j}\,\label{51}%
\end{align}
is the inverse of the $\Omega$-matrix. At this point, we assume that the
background field does not break parity symmetry. As in the Standard Model for
Particle Physics, we state that the parity symmetry breaking only occurs in
the matter sector, which is not the case studied here. Moreover, we ought to
use the Fierz rearrangement technique with the aim of forming bilinear terms
only with $\Psi$ and leaving $\Lambda_{0}^{a}$ free in eq. \eqref{47}. Also,
after following such steps, we consider the situation such that $\bar{\Psi
}\Psi$ is the only background fermion bilinear term present in our model:
\begin{align}
\bar{\Psi}\gamma_{5}\Psi &  =  0\,,\nonumber\\
\bar{\Psi}\gamma_{\mu}\gamma_{5}\Psi &  =  0\,.\label{52}%
\end{align}
Then, after applying the set of conditions \eqref{52} into the expression
\eqref{50}, the latter turns out to be written as
\begin{align}
\bigg(\omega\gamma^{0}-2M_{1}+2iM_{2}\gamma_{5}+\frac{1}{2}R^{\mu}\gamma_{\mu
}\gamma_{5}\bigg)\Lambda_{0}^{a}  &  =  2\,\omega\Omega_{ij}^{-1}%
\bigg[\frac{1}{4}\delta_{ij}\bigg(\bar{\Psi}\Psi\bigg)+\frac{i}{2}%
\bigg(\bar{\Psi}\Psi\bigg)\bigg(\Sigma_{ij}\bigg)\bigg]\Lambda_{0}%
^{a}\,.\label{53}%
\end{align}
Here, we replace eq. \eqref{51} into eq. \eqref{53} which yields the following
expression
\begin{align}
&   \bigg[\omega\gamma^{0}-2M_{1}+2iM_{2}\gamma_{5}+\frac{1}{2}R^{\mu}%
\gamma_{\mu}\gamma_{5}-\frac{1}{2}\bigg(\bar{\Psi}\Psi\bigg)\frac{1}%
{\omega^{2}-v^{2}}\bigg(3\omega^{2}-v^{2}\bigg)\nonumber\\
& - \bigg(\bar{\Psi}\Psi\bigg)\frac{\omega}{\omega^{2}-v^{2}}\bigg(\varepsilon
_{ijk}v_{k}\Sigma_{ij}\bigg)\bigg]\Lambda_{0}^{a}\nonumber\\
&  =  0\,.\label{54}%
\end{align}
From now on, it would be interesting for us make a suitable change of
variables that is defined as
\begin{align}
\mu &  \equiv \frac{1}{2}\bigg(\bar{\Psi}\Psi\bigg)\frac{1}{\omega^{2}-v^{2}%
}\bigg(3\omega^{2}-v^{2}\bigg)\label{55}%
\end{align}
and
\begin{align}
\xi_{ij}  &  =  -\bigg(\bar{\Psi}\Psi\bigg)\frac{\omega}{\omega^{2}-v^{2}%
}\bigg(\varepsilon_{ijk}v_{k}\bigg)\,.\label{56}%
\end{align}
Moreover, we shall adopt other special conditions, for instance, by taking
$F_{background}=0$, we have, from the set \eqref{130} only the terms $M_{1} =
\frac{1}{4}\bar{\Psi}\Psi$ and $\vec{R} = \vec{v}$; therefore, one may notice that
$|\vec{v}|$ and the fermionic bilinear term $(\bar{\Psi}\Psi)$ compete to
generate the gaugino mass.At this point, we must emphasize that $M_{2}$ has been written only in terms of Im(F) due to the choice that we have made in the  set of equations cast in \eqref{52}, and, as a consequence, it vanishes whenever $F_{background}=0$.
As a result, only $\mu$ survives in this very special
circumstance, since the higher-order parameters, ($\mathcal{O}(2)$) in $v$, from LSV
are too small and can be neglected, namely
\begin{equation}
\mu = \frac{3}{2}\bigg(\bar{\Psi}\Psi\bigg)\frac{\omega^{2}}%
{\omega^{2}-v^{2}}=\frac{3}{2}\bigg(\bar{\Psi}\Psi\bigg)\bigg(1+\frac{v^{2}%
}{\omega^{2}-v^{2}}\bigg)=  \frac{3}{2}\bigg(\bar{\Psi}\Psi\bigg)+\mathcal{O}(2)\,.\label{57}%
\end{equation}
We recall the reader that, from eq.\eqref{56} the term $(\bar{\Psi}\Psi)v$ is of order $\mathcal{O}(2)$ in the Lorentz-breaking parameter; as a consequence, we decided do not consider such a term in the next steps. One then replaces eq. \eqref{57} into eq. \eqref{54} to rewrite the latter in
a more compact way and, as a result, we have the following matrix dubbed
$\Delta^{ab}_{\Lambda}$ which is the well-expected dispersion relation for the gauginos (see Appendix \ref{B} for more details),
i.e.,
\begin{align}
\Delta_{\Lambda}  &  \equiv \det\bigg(\omega\gamma^{0}+\bigg(\bar{\Psi}%
\Psi\bigg)+\frac{1}{2}v^{i}\gamma_{i}\gamma_{5}\bigg) = 0\,,
\label{dispersion}
\end{align}
where $\mu-2M_{1} = \bar{\Psi}\Psi$ and it does not matter whether
set down $v^{i}$ or $v_{i}$ since we are computing the
quantities in Euclidean space. As $\omega$ corresponds to the rest mass
whenever ($\vec{k}=0$), and the sign of the mass in Dirac's equation is irrelevant
due to a possible redefinition of the field through a $\gamma_{5}$-matrix,
i.e. $-m\bar{\lambda}\lambda= +m(\overline{\gamma_{5}\lambda})(\gamma
_{5}\lambda)$, we have only two independent solutions, which yield the
following result for the gaugino mass:
\begin{align}
m_{\Lambda}  &  =  \bar{\Psi}\Psi\pm\frac{1}{2}|\vec{v}%
|\,.\label{gauginosmass}%
\end{align}
To compute the gauge boson mass, we can invert the procedure done for
the gaugino, i.e. we replace $\Lambda_{0}^{a}$ written in terms of the
Yang-Mills electric and magnetic fields into the Amp\'{e}re-Maxwell law for
Yang-Mills \eqref{43}. Then, after some calculations, one finds the dispersion
relation for the gauge boson defined as
\begin{align}
\Delta_{A}  & \equiv \det(\omega\delta_{ij}+i\epsilon_{ijk}v_{k}) = \omega(\omega^{2}-|\vec
{v}|^{2})\,,
\end{align}
that yields the following result for its mass
\begin{align}
m_{A}  &  =  |\vec{v}|\,.\label{gauge fieldmass}%
\end{align}
As one sees from eq. \eqref{gauge fieldmass}, the gauge boson mass does not
depend neither on the fermionic bilinears $(M_{1},M_{2})$ nor on the scalar
background, F; thus, the unique dependence in the expression above is on the
background vector background, $\vec{R}=\vec{v}$, and this is the reason why we
have considered, as a particular case, only a $\Psi$-fermion bilinear as being
non-trivial. Then, a massless gauge boson that appears whenever $|\vec{v}|=0$
does not imply massless gauginos, since we still have here the contribution
coming from the fermionic bilinear term $\bar{\Psi}\Psi$ in
\eqref{gauginosmass}. Indeed, LSV induces a SUSY breaking which becomes
manifest with the splitting above of the gauge boson and gaugino masses. In
addition, we notice a way to observe a potential path to outline the
conversion of massive gauginos into gauge bosons, as induced by the background
fermion that mixes them up.

The Dirac equation for the gauginos field eq. \eqref{gauginos1} must be work out at
this moment, in other words, one may develop the computation of the Gordon decomposition for the
gaugino axial current. Let us recall that, by virtue of the Majorana nature of
the gaugino, a vector-type current (diagonal in the gauge group index) is
identically vanishing. Thus, by iterating the Dirac equation and by making
use of specific properties of Majorana-fermion bilinears and a number of
special relations of the Clifford algebra, we arrive at the two equations cast
below:
\begin{align}
\bar{\Lambda}^{a}\gamma^{\nu}\gamma_{5}\Lambda^{a}  & = - \frac{i}{2M_{1}}%
\bar{\Lambda}^{a}\gamma_{5} \mathcal{D}^{\nu}\Lambda^{a} - \frac{i}{2M_{1}%
}\epsilon^{\nu\mu\kappa\lambda}\bar{\Lambda}^{a} \Sigma_{\mu\kappa}
\mathcal{D}_{\lambda}\Lambda^{a}+ \frac{\sqrt{2}}{2M_{1}}\bar{\Psi}\gamma
_{\mu}\gamma_{5} \Lambda^{a} \tilde{F}^{\nu\mu\,a}\nonumber\\
& - \frac{1}{4M_{1}}R^{\nu}\bar{\Lambda}^{a}\Lambda^{a}\label{eq_gord1}%
\end{align}
and
\begin{align}
\bar{\Lambda}^{a} \mathcal{D}^{\nu}\Lambda^{a} - 2i \bar{\Lambda}^{a}
\Sigma^{\nu\mu} \mathcal{D}_{\mu}\Lambda^{a} + 2M_{2} \bar{\Lambda}^{a}%
\gamma^{\nu}\gamma_{5}\Lambda^{a} -\frac{i}{2}R^{\nu}\bar{\Lambda}^{a}%
\gamma_{5}\Lambda^{a} - 2\sqrt{2}\bar{\Psi}\gamma_{\mu}\gamma_{5}\Lambda^{a}
F^{\nu\mu\,a} = 0\label{eq_gord2}%
\end{align}
where we recall that the gauge covariant derivative acting on the gaugino
reads as follows:
\begin{equation}
\mathcal{D}_{\mu}\Lambda^{a} = (\delta^{ac}\partial_{\mu}- g f^{abc}A_{\mu
}^{b} )\Lambda^{c}\,.
\end{equation}

Note that the term in $\tilde{F}^{\mu\nu\, a}$ appears in the chiral
current decomposition eq. (\ref{eq_gord1}) and incorporates the term $\bar{\Psi}%
\vec{\gamma}\gamma_{5} \Lambda^{a} \cdot\vec{B}^{a}$, which shows a sort of
transition Yang-Mills magnetic coupling between the gaugino and the background
fermion, $\Psi$. If we go back to the Abelian scenario, such a term would be
pointing to a Pauli-type magnetic interaction, present by virtue of the
photino mass, which indicates that though both the gaugino and the background
fermion are electrically neutral, they develop a transition magnetic dipole
moment, very similar to what happens to the transition magnetic moments of
massive Majorana neutrinos.

Finally, we draw attention to the new term that appears as a result of the coupling
with the background field. It is a spin-orbit term. The traditional spin-orbit
interaction generated novel materials known as topological materials in which
the quantum spin Hall effect \cite{hasan2010colloquium} arises. The type of Lorentz symmetry violation that we worked on in our article begins to be investigated in
Quantum Mechanics \cite{belich2015spin, bakke2014rashba} and serves as a guide
for investigations in a context of physics beyond the Standard Model with
spontaneous SUSY violation.


\section{Concluding Comments}

\label{V}

Initially, we have reviewed the pure super-Yang-Mills action to set up our notations and conventions adopted throughout this paper. After that, we followed our viewpoint such that LSV induces the breaking of SUSY. To do that, we have proposed an $\mathcal{N}=1$ supersymmetric extension for a non-Abelian gauge model with the Yang-Mills version of the
Carroll-Field-Jackiw topological term. As a consequence, we verify that the non-trivial SUSY transformation of the background fermion, $\psi$, signals SUSY breaking. Following that, we have rewritten the action \eqref{SuCS} in terms of Majorana $4$-component spinors, then, after removing the auxiliary field, $D$, we attained a form of the action where the mixing between the gauge boson and gaugino appears thanks to the fermionic component of the SUSY background.

Following that, we have focused on computing the gauge field and gaugino dispersion relations to obtain their respective masses. To get the latter, we have chosen to work with the linearized regime of their equations of motion in terms of the Yang-Mills electric and magnetic fields. Thus, due
to the linearization, we have replaced these fields with their plane wave solutions. We could note that there is a competition between LSV and SUSY effects in the final expression for the gaugino mass \eqref{gauginosmass}.
Also, we have been able to find a topological mass for the gauge field written in terms of the vector background field, $\vec{v}$, as a result of the LSV phenomenon. Here, we ought to say that we consider such mass as a feasible small correction when one compares it with the gauge field screening mass observed in lattice simulations and investigated in some non-perturbative models; for that, we suggest the reader to consult the works \cite{Cucchieri:2007md,Dudal:2008sp,PhysRevD.84.045018,10.21468/SciPostPhys.10.2.035}%
.

Finally, the preliminary results reported in this contribution point to further potential investigations to be pursued. First, as an immediate study, it would be worthwhile to compute the expression for the probability of conversion of massive gauginos into gauge bosons as a consequence of their
mixing. With the present limits on the LSV parameters involved, we shall be able to estimate this effect. Next, an issue to be pursued can be the attainment of the gaugino decay into a gauge boson pair as a consequence of the gautino-gauge boson-gauge boson coupled to the background vertex with gauge coupling constant. As a follow-up, we intend to quantize the SYM-CFJ action, focusing on its renormalization and the investigation of the contribution of the $\Psi$ and $v$ partners to both the gauge field and gaugino (background-)induced masses at the one-loop level. Also, by introducing the quark sector and considering the special case of SUSY QCD in presence of LSV, attainment of the glueball spectrum and comparison with lattice calculations could place new limits on the LSV parameters imposed by glueball masses. Any progress in any of these directions shall be reported elsewhere in forthcoming papers.


\section*{Acknowledgments}
The authors would like to thank L.P.R. Ospedal for reading the manuscript. R. C. Terin is supported by the National Council for Scientific and Technological Development from Ministry for Science, Technology, and Innovations (CNPq/MCTI) through the junior postdoctoral fellowship program (PDJ), Grant - 151397/2020-1. W. Spalenza, Thanks Scuola Internazionale Superiore di Studi Avanzati (SISSA) and Loriano Bonora, for hospitality during Postdoc.

\begin{appendix}
\section{Superspace Formalism and Conventions}\label{A}
In this appendix, we will establish our notations and conventions.\\
\noindent
\textbf{Units}: $\hbar=c=1$.\\
\noindent
\textbf{Minkowski metric}: $\eta_{\mu\nu}=diag(+,-,-,-)$.\\
\noindent {\bf Indices notations}:
\begin{center}
\begin{tabular}{ll}
$\bullet$&The Lorentz indices: $\mu,\nu,\rho,\sigma,\lambda, ... \in\{0,1,2,3\}$\,;$\phantom{\Bigl|}$\\
$\bullet$&The Spinor indices: $\alpha,\beta,\gamma,\delta,\eta, ... \in\{1,2,3,4\}$\,;$\phantom{\Bigl|}$\\
$\bullet$&The $SU(N)$ group indices: $a,b,c,d,e, ... \in\{1,\dots,N^{2}-1\}$\,.$\phantom{\Bigl|}$\\
\end{tabular}
\end{center}
Now, let us present some useful expressions that are currently necessary when carrying out spinorial calculations in the Weyl's representation. First, the relativistic matrices $%
\sigma ^{\mu }=\left( \sigma _{\alpha \dot{\alpha}}^{\mu }\right) $ and $%
\bar{\sigma}^{\mu }=\left( \bar{\sigma}_{\dot{\alpha}\alpha }^{\mu
}\right) $ are defined as: $\sigma ^{\mu }=(1,\sigma
^{i}),\,\,\bar{\sigma}^{\mu
}=(1,-\sigma ^{i}),\,\,$where $\sigma ^{i}$ are the Pauli matrices and $i$ runs over ($1, 2, 3$). Next, the $SO(1,3)$ group generators are represented by the following matrices
\begin{equation}
\sigma ^{\mu \nu }=\frac{i}{4}(\sigma ^{\mu }\bar{\sigma}^{\nu
}-\sigma
^{\nu }\bar{\sigma}^{\mu }),\,\,\,\,\,\,\,\,\,\,\bar{\sigma}^{\mu \nu }=%
\frac{i}{4}(\bar{\sigma}^{\mu }\sigma ^{\nu }-\bar{\sigma}^{\nu
}\sigma ^{\mu })\,.
\end{equation}
An interesting relation involving $\sigma $ matrices is
\begin{equation}
\sigma _{\alpha \dot{\alpha}}^{\mu }\bar{\sigma}^{\nu
\,\dot{\alpha}\alpha
}\sigma _{\beta \dot{\beta}}^{\kappa }\bar{\sigma}^{\lambda \,\dot{\beta}%
\beta }=2(\eta ^{\mu \nu }\eta ^{\kappa \lambda }-\eta ^{\mu
\kappa }\eta ^{\nu \lambda }+\eta ^{\mu \lambda }\eta ^{\nu \kappa
}-i\varepsilon ^{\mu \nu \kappa \lambda }),
\end{equation}
where $\varepsilon ^{0123}=-\varepsilon _{0123}=1.$ In addition,
the Grassmannian coordinates $\theta $ and $\bar{\theta}$ have
their indices raised and lowered as
\begin{equation}
\theta ^{\alpha }=\varepsilon ^{\alpha \beta }\theta _{\beta
},\,\,\,\,\,\,\,\,\,\theta _{\alpha }=\varepsilon _{\alpha \beta
}\theta
^{\beta },\,\,\,\,\,\,\,\,\,\bar{\theta}^{\dot{\alpha}}=\varepsilon ^{\dot{%
\alpha}\dot{\beta}}\bar{\theta}_{\dot{\beta}},\,\,\,\,\,\,\,\,\,\bar{\theta}%
_{\dot{\alpha}}=\varepsilon _{\dot{\alpha}\dot{\beta}}\bar{\theta}^{\dot{%
\beta}},
\end{equation}
with, $
\varepsilon ^{12}=\varepsilon _{21}=\varepsilon ^{\dot{1}\dot{2}%
}=\varepsilon _{\dot{2}\dot{1}}=1$ and $\varepsilon ^{\alpha \beta
}=-\varepsilon ^{\beta \alpha }$. Furthermore, the square spinor definitions are
\begin{equation}
\theta ^{2}=\theta ^{\alpha }\theta _{\alpha }=-2\theta ^{1}\theta
^{2},\,\,\,\,\,\,\,\,\,\,\,\,\,\,\bar{\theta}^{2}=\bar{\theta}_{\dot{\alpha}}%
\bar{\theta}^{\dot{\alpha}}=2\bar{\theta}_{\dot{1}}\bar{\theta}_{\dot{2}}\,,
\end{equation}
and
\begin{eqnarray}
\theta ^{\alpha }\theta ^{\beta } &=&-\frac{1}{2}\varepsilon
^{\alpha \beta
}\theta ^{2},\,\,\,\,\,\,\theta _{\alpha }\theta _{\beta }=\frac{1}{2}%
\varepsilon _{\alpha \beta }\theta ^{2}, \\
\bar{\theta}^{\dot{\alpha}}\bar{\theta}^{\dot{\beta}} &=&\frac{1}{2}%
\varepsilon ^{\dot{\alpha}\dot{\beta}}\bar{\theta}^{2},\,\,\,\,\,\,\bar{%
\theta}_{\dot{\alpha}}\bar{\theta}_{\dot{\beta}}=-\frac{1}{2}\varepsilon _{%
\dot{\alpha}\dot{\beta}}\bar{\theta}^{2}\,,
\end{eqnarray}
where $\varepsilon _{\alpha \beta }\varepsilon ^{\gamma \delta
}=\delta _{\alpha }^{\,\,\,\,\,\gamma }\delta _{\beta
}^{\,\,\,\,\,\delta }-\delta _{\alpha }^{\,\,\,\,\,\delta }\delta
_{\beta }^{\,\,\,\,\,\gamma },$ the same for conjugate$.$ Other useful relations are
\begin{equation}
\sigma _{\alpha \dot{\alpha}}^{\mu
}\bar{\theta}^{\dot{\alpha}}(\theta
\sigma ^{\nu }\bar{\theta})=\frac{1}{2}\eta ^{\mu \nu }\theta _{\alpha }\bar{%
\theta}^{2},\,\,\,\,\,\,\,\,\,\,\,\,\,\,\,(\theta \sigma ^{\mu }\bar{\theta}%
)(\theta \sigma ^{\nu }\bar{\theta})=\frac{1}{2}\eta ^{\mu \nu }\theta ^{2}%
\bar{\theta}^{2}.
\end{equation}
The superalgebra is defined as follows:
\begin{equation}
\{Q^{I},Q^{J}\}=\{\bar{Q}^{I},\bar{Q}^{J}\}=0,\,\,\,\,\,\,\,\,\,\,\,\{Q_{%
\alpha }^{I},\bar{Q}_{\dot{\alpha}}^{J}\}=\delta ^{IJ}\sigma _{\alpha \dot{%
\alpha}}^{\mu }\,p_{\mu }\,,
\end{equation}
with $I, \,J=1,2$. The covariant derivatives are defined by
\begin{equation}
\mathcal{D}_{\alpha }=\partial _{\alpha }+i\sigma _{\alpha
\dot{\alpha}}^{\mu }\theta
^{\dot{\alpha}}\partial _{\mu },\,\,\,\,\,\,\,\,\,\,\,\,\,\,\,\,\,\,\,\bar{\mathcal{D}}%
_{\dot{\alpha}}=-\partial _{\dot{\alpha}}-i\theta ^{\alpha }\sigma
_{\alpha \dot{\alpha}}^{\mu }\partial _{\mu }, \label{cov_deriv_alpha}
\end{equation}
where both anti-commutes with $Q$ and $\bar{Q},$ and the self
anti-commutation between them is established as
\begin{equation}
\{\mathcal{D}_{\alpha },\bar{\mathcal{D}}_{\dot{\alpha}}\}=-2i\sigma _{\alpha \dot{\alpha}
}^{\mu }\partial _{\mu }.
\end{equation}
Finally, if one wishes to describe a supersymmetric action, one needs to define the rules for the integration over the Grassmannian coordinates. In the case $D= 1+3$, one has the following relations:
\begin{equation}
d^{2}\theta =\frac{1}{4}d\theta ^{\alpha }d\theta ^{\beta
}\varepsilon
_{\beta \alpha },\,\,\,\,d^{2}\bar{\theta}=\frac{1}{4}d\bar{\theta}_{\dot{%
\alpha}}d\bar{\theta}_{\dot{\beta}}\varepsilon ^{\dot{\beta}\dot{\alpha}%
},\,\,\,\,\,\,\,\,d^{4}\theta =d^{2}\theta d^{2}\bar{\theta},\nonumber
\end{equation}
\begin{equation}
\int d^{2}\bar{\theta}\left( \bar{\theta}^{2}\right)
=\int d^{2}\theta \left( \theta ^{2}\right) =1,\nonumber
\end{equation}
\begin{equation}
\int d^{2}\bar{\theta}\left( \bar{\theta}_{\dot{\alpha}}\right)
=\int d^{2}\theta \left( \theta ^{\alpha}\right) =0,
\end{equation}
which yield the expressions below
\begin{equation}
\delta^{2}\left(
\theta \right) =\theta ^{2},\,\,\,\,\,\,\delta ^{2}\left( \bar{\theta}%
\right) =\bar{\theta}^{2}.
\end{equation}
With them, a superfield integral is written as
\begin{equation}
\int d^{4}xd^{2}\theta
d^{2}\bar{\theta}\,\Phi_{\theta^{2}\bar{\theta}^{2}}.
\end{equation}
Thus, the free Wess-Zumino action in terms of the superfields is
\begin{equation}
{\cal S} =\int d^{4}xd^{2}\theta
d^{2}\bar{\theta}\,\bar{\Phi}(x,\theta,\bar{\theta})\Phi(x,\theta,\bar{\theta}).
\end{equation}

\section{Computation of the gaugino mass spectrum}\label{B}

In order to compute the gaugino masses obtained in eq.\eqref{gauginosmass} we must first present again the dispersion relation expression showed in eq.\eqref{dispersion} 
\begin{eqnarray}
\Delta_{\Lambda} &=& \det\bigg(\omega\gamma^{0}\delta^{ab}+\bigg(\bar{\Psi}\Psi\bigg)\delta^{ab}+\frac{1}{2}\delta^{ab}v_{i}\gamma_{i}\gamma_{5}\bigg) = 0\,.
\end{eqnarray}
Thus, one has to follow the steps below
\begin{eqnarray}
\bigg[(\omega+\bar{\Psi}\Psi)^{2}-\frac{1}{4}v^{2}\bigg]\bigg[(\omega-\bar{\Psi}\Psi)^{2}-\frac{1}{4}v^{2}\bigg] & = & 0\,,
\end{eqnarray}
where $\omega$ corresponds to the mass ($\vec{k}=0$):
\begin{eqnarray}
|\omega+\bar{\Psi}\Psi| & = & \frac{1}{2}|\vec{v}|
\end{eqnarray}
and 
\begin{eqnarray}
|\omega-\bar{\Psi}\Psi| & = & \frac{1}{2}|\vec{v}|\,.
\end{eqnarray}
An important remark is that $\bar{\Psi}\Psi$ is real, but it is not necessarily positive-definite. Therefore, let us analyze the possibilities:
\begin{itemize}
\item For $\bigg(\omega+\bar{\Psi}\Psi\bigg)\geq0$: 
\begin{eqnarray}
\omega_{1}+\bar{\Psi}\Psi & = & \frac{1}{2}|\vec{v}|\nonumber \\
\omega_{1} & = & -\bar{\Psi}\Psi+\frac{1}{2}|\vec{v}|\,.
\label{B1}
\end{eqnarray}
\item If $\bigg(\omega+\bar{\Psi}\Psi\bigg)<0$, then:
\begin{eqnarray}
\omega_{2}+\bar{\Psi}\Psi & = & -\frac{1}{2}|\vec{v}|\nonumber \\
\omega_{2} & = & -\bar{\Psi}\Psi-  \frac{1}{2}|\vec{v}|\,.
\label{B2}
\end{eqnarray}
\item For $\bigg(\omega-\bar{\Psi}\Psi\bigg)\geq0$:
\begin{eqnarray}
\omega_{3}+\bar{\Psi}\Psi & = & \frac{1}{2}|\vec{v}|\nonumber \\
\omega_{3} & =&-\bar{\Psi}\Psi+ \frac{1}{2}|\vec{v}|\,.
\label{B3}
\end{eqnarray}
\item If $\bigg(\omega-\bar{\Psi}\Psi\bigg)<0$, then:
\begin{eqnarray}
\omega_{4}+\bar{\Psi}\Psi & = & -\frac{1}{2}|\vec{v}|\nonumber \\
\omega_{4} & = &-\bar{\Psi}\Psi- \frac{1}{2}|\vec{v}|\,.
\label{B4}
\end{eqnarray}
\end{itemize}
From the set above, i.e., eqs \eqref{B1} - \eqref{B4}, one has the following results
\begin{eqnarray}
\omega_{4} & = & -\omega_{1}\,,\qquad
\omega_{2} =  -\omega_{3}\,.
\end{eqnarray}
From the main text, we observed that the sign of the mass in Dirac's equation is irrelevant due to a possible redefinition of the field through a $\gamma_{5}$ matrix, i.e.
\begin{eqnarray*}
-m\bar{\lambda}\lambda & = & +m(\overline{\gamma_{5}\lambda})(\gamma_{5}\lambda),
\end{eqnarray*}
we have only two independent solutions, for instance
\begin{eqnarray}
\omega_{3} & = & \bar{\Psi}\Psi+\frac{1}{2}|\vec{v}|
\end{eqnarray}
and 
\begin{eqnarray}
\omega_{4} & = & \bar{\Psi}\Psi-\frac{1}{2}|\vec{v}|\,.
\end{eqnarray}
Finally we arrive at the eq. \eqref{gauginosmass} rewritten below
\begin{eqnarray}
m_{\Lambda} & = & \bar{\Psi}\Psi\pm\frac{1}{2}|\vec{v}|\,.
\end{eqnarray}
\end{appendix}



\end{document}